\documentclass[aps,pra,twocolumn,showpacs,preprintnumbers,amsmath,amssymb,superscriptaddress]{revtex4-2}

\usepackage{graphicx}
\usepackage{epsfig}
\usepackage{amsmath}
\usepackage{amssymb}
\usepackage{textcomp}
\usepackage{dcolumn}
\usepackage{bm}
\usepackage{braket}
\usepackage{bbm}
\usepackage[usenames,dvipsnames]{xcolor}
\usepackage[colorlinks=true,citecolor=MidnightBlue,linkcolor=MidnightBlue,urlcolor=MidnightBlue]{hyperref}
\usepackage{siunitx}

\PassOptionsToPackage{numbers,sort&compress}{natbib}

\makeatletter
\g@addto@macro\normalsize{%
  \setlength\abovedisplayskip{4pt}
  \setlength\belowdisplayskip{4pt}
  \setlength\abovedisplayshortskip{4pt}
  \setlength\belowdisplayshortskip{4pt}
}
\makeatother

\renewcommand{\BibitemShut}[1]{} 
\renewcommand{\vec}[1]{\mathbf{#1}}

\begin{document}

\newcommand*{\MAINZ}{QUANTUM, Institut f\"ur Physik, Johannes Gutenberg-Universit\"at Mainz, Staudingerweg 7, 55128 Mainz, Germany}
\newcommand*{\ERLANGEN}{Institut f\"ur Optik, Information und Photonik, Friedrich-Alexander Universit\"at Erlangen-N\"urnberg, Staudtstr. 1, 91058 Erlangen, Germany}
\newcommand*{\SAOT}{Erlangen Graduate School in Advanced Optical Technologies (SAOT), Friedrich-Alexander Universit\"at Erlangen-N\"urnberg, Paul-Gordan-Str. 6, 91052 Erlangen, Germany}
\homepage{http://www.quantenbit.de}

\title{Different Types of Coherence: Young-type Interference versus Dicke Superradiance}

\author{D.~Bhatti}
\affiliation{Institut f\"{u}r Optik, Information und Photonik, Friedrich-Alexander-Universit\"{a}t Erlangen-N\"{u}rnberg (FAU), 91058 Erlangen, Germany}
\affiliation{Institute for Functional Matter and Quantum Technologies, Universität Stuttgart, 70569 Stuttgart, Germany}
\affiliation{Erlangen Graduate School in Advanced Optical Technologies (SAOT), Friedrich-Alexander-Universit\"{a}t Erlangen-N\"{u}rnberg (FAU), 91052 Erlangen, Germany}
\author{M.~Bojer}
\affiliation{Institut f\"{u}r Optik, Information und Photonik, Friedrich-Alexander-Universit\"{a}t Erlangen-N\"{u}rnberg (FAU), 91058 Erlangen, Germany}
\author{J.~von~Zanthier}
\affiliation{Institut f\"{u}r Optik, Information und Photonik, Friedrich-Alexander-Universit\"{a}t Erlangen-N\"{u}rnberg (FAU), 91058 Erlangen, Germany}
\affiliation{Erlangen Graduate School in Advanced Optical Technologies (SAOT), Friedrich-Alexander-Universit\"{a}t Erlangen-N\"{u}rnberg (FAU), 91052 Erlangen, Germany}

\date{\today}

\begin{abstract}
Dicke superradiance, i.e., the enhanced spontaneous emission of coherent radiation, is often attributed to radiation emitted by synchronized dipoles coherently oscillating in phase. At the same time, Dicke derived superradiance assuming atoms in entangled Dicke states which  do not display any dipole moment. To shed light on this apparent paradox, we study the intensity distribution arising from two identical two-level atoms prepared either in an entangled Dicke state or in a separable atomic state with non-vanishing dipole moment. We find that the two configurations produce similar far field intensity patterns, however, stemming from fundamentally distinct types of coherence: while in the second case the atoms display coherence among the individual particles leading to Young-type interference as known from classical dipoles, atoms in Dicke states possess collective coherence leading to enhanced spontaneous emission. This demonstrates that the radiation generated by synchronized dipoles and Dicke superradiance are fundamentally distinct phenomena and have to be interpreted in different ways.
\end{abstract}

\maketitle

In a seminal paper in 1954, R.~H. Dicke demonstrated that an ensemble of $N$ two-level atoms prepared in entangled symmetric Dicke states radiates spontaneous emission with enhanced intensity scaling with a factor of up to $N^{2}$ \cite{Dicke(1954)}. Since then, this phenomenon has been commonly attributed to a collection of atomic dipoles oscillating in phase, equivalent to coherently oscillating antennas \cite{Bonifacio(1975),Haroche(1982),MANDEL(1995),Scully(2009)Super} or even resonant acoustic waves in a piano \cite{MANDEL(1995)}. However, this interpretation ignores the fact that symmetric Dicke states do not carry any dipole moment \cite{Wiegner(2011)PRA}. The question is thus why there has not been established a clear distinction between the classical phenomenon of synchronized dipoles or antennas oscillating in phase and the quantum mechanical effect of Dicke superradiance.

One possible reason is that the different configurations produce similar interference effects. In particular, considering the far field intensity distribution of the emitted radiation, one finds that coherently driven atoms \cite{Eichmann(1993),DeVoe(1996),Itano(1998),Skornia(2001),Wolf(2016)}, atoms in product states \cite{Mandel(1983)}, and two-level atoms prepared in symmetric Dicke states \cite{Wiegner(2011)PRA,Oppel(2014),Wiegner(2015)} display similar intensity patterns, although potentially of different contrast. Since Young's famous double-slit experiments \cite{YOUNG(1807)}, these patterns are well-known in optics and attributed to coherent radiation, i.e., coherently oscillating dipoles or slits and sources emanating light with a fixed mutual phase relation. However, a more detailed analysis reveals that the interference patterns stemming from synchronized dipoles and Dicke superradiance are fundamentally distinct phenomena and have to be interpreted in  different manners. In this paper we will discern the different types of coherence present in the different atomic configurations, leading to a more profound understanding of the physical origin of the observed phenomena.

To quantify the collective coherence of the different configurations several coherence monotones based on distance measures could be used \cite{Baumgratz(2014),Streltsov(2015),Streltsov(2017),streltsov(2018),Bartosz(2018),Theurer(2019),Xu(2020),wu(2020),yuan(2020)}. Among them, the quantum discord has been shown to indicate quantum correlations even beyond entanglement~\cite{Modi(2012)} and, moreover, to be equivalent to basis-free quantum coherence~\cite{Yao(2015)}. We will therefore employ the quantum discord to characterize the \textit{quantum coherence} of the different configurations, as well as the well-established measure concurrence. In contrast, for states which do not display quantum coherence but nonetheless produce far field interference patterns, we will use a classical measure in the form of the global dipole moment to identify the \textit{classical coherence} of the corresponding states.

In what follows, we will investigate  two two-level atoms prepared in three different configurations, namely both atoms in a symmetric Dicke state with one excitation; both atoms in the same coherent superposition of ground and excited state; and both atoms in a Werner state. In this way, we study examples for systems possessing exclusively quantum coherence and configurations exhibiting solely classical coherence; at the same time, all three configurations produce far field interference patterns of similar form, yet with slightly varying visibilities. We will determine the specific quantum and classical coherence for each of these configurations by calculating on the one hand their quantum discord and concurrence, and on the other hand their classical dipole moment. In particular, we will see that in contrast to the widespread opinion that a strong global dipole moment causes atoms in a symmetric Dicke state to radiate in a coherent manner with  a visibility of $100\%$, the dipole moment of these states vanishes and only quantum coherence in the form of entanglement exists, leading to interference in Dicke's celebrated spontaneous emission of coherent radiation. On the contrary, we will derive that a system of two synchronously oscillating dipoles solely possesses a classical dipole moment, i.e., classical coherence, and produces an interference pattern resulting from the coherently radiated emission with a visibility $< 100\%$. Finally, by investigating the Werner state, we will establish that the visibility of the interference patterns alone does not deliver any information about the type of coherence. In total, we show that the quantum mechanical phenomenon of Dicke superradiance and the intensity distribution stemming from coherently radiating dipoles are fundamentally distinct phenomena and have to be explained from inherently different perspectives~\cite{Agarwal(1974)}.

The paper is organized as follows: in Sec.~\ref{sec:SystemMeasures} we introduce the system of two two-level atoms together with the corresponding density matrix and calculate the spatial intensity distribution produced by the system in the far field; the latter will reveal which elements of the density matrix lead to interferences and to a modulated intensity profile; further, we will introduce the quantum discord together with the concurrence as a measure for quantum coherence, and the classical dipole moment as a measure for classical coherence. In Sec.~\ref{sec:SymDickeState}, we investigate the two-atom system prepared in a symmetric Dicke state with one excitation, producing an intensity pattern with a visibility of 100$\%$, and show that this state only possesses quantum coherence. This is in contrast to a system of two identical synchronized atomic dipoles which turns out to possess only classical coherence, what is discussed in Sec.~\ref{sec:Dipoles}; we will also find that this classical system can not produce an interference pattern with a visibility of 100$\%$. In Sec.~\ref{sec:WernerState}, we prove that the visibility of the interference pattern is not a measure for the type of coherence, since two two-level atoms prepared in a Werner state produce an interference pattern of arbitrary visibility although the state possesses only quantum coherence. In Sec.~\ref{sec:conclusion} we finally conclude.

\begin{figure}[t]
	\centering
		\includegraphics[width=0.8\columnwidth]{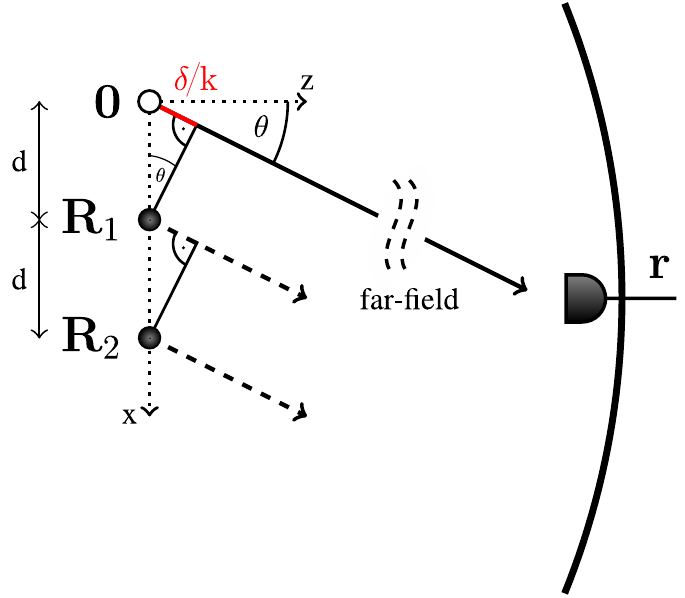}
	\caption{Investigated setup: two two-level atoms, located along the x-axis at positions $\vec{R}_{1}$ and $\vec{R}_{2}$ separated by a distance $d\gg \lambda$, are prepared in different states, while a detector measures the intensity distribution $G^{(1)}(\vec{r})$ at position $\vec{r}$ in the far field. The dipole moment of the two atoms is assumed to be oriented perpendicular to the indicated plane.
		\vspace{0.3cm}}
	\label{fig:Setup_21}
\end{figure}

\section{System under investigation}
\label{sec:SystemMeasures}
In this section we introduce the system consisting of two two-level atoms prepared in different initial states and discuss the spatial intensity pattern produced by the atoms in the far field. The two atoms are assumed to be placed along the x-axis at positions $\vec{R}_{1}$ and $\vec{R}_{2}$, separated by a distance $d\gg \lambda$ such that the dipole-dipole interaction or any other kind of direct interaction can be neglected (see Fig.~\ref{fig:Setup_21}). Denoting with $\ket{e}_{l}$ ($\ket{g}_{l}$) the excited (ground) state of the $l$th atom, $l\in\lbrace 1,2\rbrace$, we can write an arbitrary state of the two atoms in the form of the following Hermitian positive semi-definite density matrix
\begin{align}
	\hat{\rho} = \bordermatrix{
			    ~ & gg & eg & ge & ee \cr
		gg & a_{\ket{gg}\bra{gg}} & a_{\ket{gg}\bra{eg}} & a_{\ket{gg}\bra{ge}} & a_{\ket{gg}\bra{ee}} \cr
    eg & a_{\ket{eg}\bra{gg}} & a_{\ket{eg}\bra{eg}} & a_{\ket{eg}\bra{ge}} & a_{\ket{eg}\bra{ee}} \cr
		ge & a_{\ket{ge}\bra{gg}} & a_{\ket{ge}\bra{eg}} & a_{\ket{ge}\bra{ge}} & a_{\ket{ge}\bra{ee}} \cr
		ee & a_{\ket{ee}\bra{gg}} & a_{\ket{ee}\bra{eg}} & a_{\ket{ee}\bra{ge}} & a_{\ket{ee}\bra{ee}}  } ,
		\label{eq:ArbMatrix}
\end{align}
where $\ket{ij}=\ket{i}_{1} \otimes \ket{j}_{2}$, with $i,j \in \left\{e,g\right\}$.
To obtain a normalized state, the condition
\begin{align}
	\text{Tr}[\hat{\rho}] = 1 \, 
\end{align}
holds, where Tr$[f]$ denotes the trace of the matrix $f$.
Note that for simplicity we choose all matrix elements $a_{\ket{ij}\bra{kl}} \in \mathbb{R}$, $i,j,k,l\in\left\{e,g\right\}$, so that the density matrix of Eq.~(\ref{eq:ArbMatrix}) becomes symmetric, i.e., $a_{\ket{ij}\bra{kl}}=a_{\ket{kl}\bra{ij}}$.

The spatial intensity distribution in the far field  produced by the atomic state $\hat{\rho}$ at the position $\vec{r}$ is given by the first-order intensity correlation function \cite{Glauber(1963)Coherent}
\begin{equation}
	G^{(1)}_{\hat{\rho}}(\vec{r})=\left< \hat{E}^{(-)}(\vec{r}) \hat{E}^{(+)}(\vec{r}) \right>_{\hat{\rho}} \, .
\label{eq:G1Def}
\end{equation}
Here, the positive and negative parts of the electric field operator are given by \cite{Wiegner(2015)}
\begin{equation}
\left[ \hat{E}^{(-)}(\vec{r}) \right]^{\dagger}	= \hat{E}^{(+)}(\vec{r}) \sim \sum_{l=1}^{2} e^{-i l \delta} \hat{s}_{-}^{(l)}
\, ,
\label{eq:Efield}
\end{equation}
where $\hat{s}_{-}^{(l)} = \ket{g}_{l} \bra{e}$ denotes the quantum mechanical lowering operator of the $l$th atom. Note that in Eq.~\eqref{eq:Efield} we assume that the dipole moment of the two atoms $\vec{d}_{eg}=\bra{e}\hat{\vec{d}} \ket{g}_{l}=\vec{d}_{ge}^*$, $l\in \lbrace 1,2\rbrace$, is oriented perpendicular to the observation plane; furthermore, the relative optical phase accumulated by a photon travelling from source $l$ to the detector at $\vec{r}$ with respect to a photon traveling from the origin is given by (see Fig.~\ref{fig:Setup_21}) 
\begin{equation}
l \delta = lk d \sin(\theta) \, .
\end{equation}

Calculating the intensity distribution by use of Eq.~(\ref{eq:Efield}), we find
\begin{align}
	G^{(1)}_{\hat{\rho}}(\vec{r}) = & \sum_{l=1}^{2} \left< \hat{s}_{+}^{(l)}\hat{s}_{-}^{(l)} \right>_{\hat{\rho}} + \sum_{l_{1}\neq l_{2}=1}^{2} e^{i  \delta (l_{1}-l_{2})} \left< \hat{s}_{+}^{(l_{1})}\hat{s}_{-}^{(l_{2})} \right>_{\hat{\rho}} \, .
\label{eq:G1Contributions}
\end{align}
Eq.~(\ref{eq:G1Contributions}) displays a constant term and an interference term, where only the latter depends on the detector position $\delta$. 
To see which elements of the matrix $\hat{\rho}$ of Eq.~(\ref{eq:ArbMatrix}) contribute to the intensity distribution of Eq.~(\ref{eq:G1Contributions}), we explicitly calculate the two terms. In this way we find 
\begin{align}
	\sum_{l=1}^{2} \left< \hat{s}_{+}^{(l)}\hat{s}_{-}^{(l)} \right>_{\hat{\rho}} & = \left< \hat{s}_{+}^{(1)}\hat{s}_{-}^{(1)} \right>_{\hat{\rho}} + \left< \hat{s}_{+}^{(2)}\hat{s}_{-}^{(2)} \right>_{\hat{\rho}} \nonumber \\
	&= a_{\ket{eg}\bra{eg}} + a_{\ket{ge}\bra{ge}} + 2a_{\ket{ee}\bra{ee}} \, ,
\label{eq:ConstantTerm}
\end{align}
for the constant term, and 
\begin{align}
	& \sum_{l_{1}\neq l_{2}=1}^{2} e^{i  \delta (l_{1}-l_{2})} \left< \hat{s}_{+}^{(l_{1})}\hat{s}_{-}^{(l_{2})} \right>_{\hat{\rho}} \nonumber \\
	& = e^{-i  \delta} \left< \hat{s}_{+}^{(1)}\hat{s}_{-}^{(2)} \right>_{\hat{\rho}} + e^{i  \delta} \left< \hat{s}_{+}^{(2)}\hat{s}_{-}^{(1)} \right>_{\hat{\rho}}  \nonumber \\[1mm]
		&= e^{-i  \delta} a_{\ket{ge}\bra{eg}} + e^{i  \delta} a_{\ket{eg}\bra{ge}} = 2 a_{\ket{ge}\bra{eg}} \cos(\delta) \, ,
\label{eq:InterferenceTerm}
\end{align}
for the interference term, where we made use of the fact that $\hat{\rho}$ is symmetric. From Eqs.~(\ref{eq:ConstantTerm}) and (\ref{eq:InterferenceTerm}) one can see that the intensity of Eq.~(\ref{eq:G1Contributions}) is of the form
\begin{align}
	G^{(1)}_{\hat{\rho}}(\vec{r}) \propto 1 + \mathcal{V}_{\rho} \cos(\delta) \, ,
\end{align}
where the visibility $\mathcal{V}_{\rho}$ of the modulation is given by
\begin{align}
	\mathcal{V}_{\rho}= \frac{2 a_{\ket{ge}\bra{eg}}}{a_{\ket{eg}\bra{eg}} + a_{\ket{ge}\bra{ge}} + 2 a_{\ket{ee}\bra{ee}}} \, .
\label{eq:Visibility}
\end{align}

Note that for two identical sources the visibility of the interference pattern is equal to the modulus of the complex degree of coherence, i.e., the normalized first-order Glauber amplitude correlation function \cite{Glauber(1963)Coherent,MANDEL(1995),LOUDON(2000),AGARWAL(2013)}.
Further, we see from Eq.~(\ref{eq:Visibility}) that the visibility disappears when the non-diagonal elements $a_{\ket{ge}\bra{eg}}$ and $a_{\ket{eg}\bra{ge}}$ in $\hat{\rho}$ vanish. These elements are known as the cross-correlations of $\hat{\rho}$ and describe the possibility that interferences \textit{between} the two atoms occur. Note further that in total Eq.~(\ref{eq:Visibility}) depends on five matrix elements, namely the four central elements, with a single excitation in each of the two two-level atoms, i.e., $\ket{i}\bra{j}$ with $i,j\in\left\{eg,ge\right\}$, and the fully excited element, i.e., $\ket{ee}\bra{ee}$ [see Eq.~(\ref{eq:ArbMatrix})]. 

In the next three chapters, we aim to understand how the system of two two-level atoms prepared in different states can produce an interference pattern of the intensity in the far field. To that aim, we investigate as witness of interference the visibility $\mathcal{V}_{\rho}$ [s. Eq.~(\ref{eq:Visibility})], as well as two different types of coherence displayed by the system, i.e., the classical coherence assessed by the total atomic dipole moment of the system, as well as the quantum coherence determined by the quantum discord and the concurrence.

The total dipole moment operator of two identical atoms can be written in the form \cite{MANDEL(1995)}
\begin{equation}
	\hat{\vec{D}} =  \vec{d}_{eg} \sum_{l=1}^{2} \hat{s}_{+}^{(l)} + \vec{d}_{ge} \sum_{l=1}^{2} \hat{s}_{-}^{(l)}\, .
\label{eq:DefD}
\end{equation}
Testing $\hat{\vec{D}}$ on the general state $\hat{\rho}$ of Eq.~(\ref{eq:ArbMatrix}) reveals that the following matrix elements are involved in the total dipole moment 
\begin{align}
	\left< \hat{\vec{D}} \right>_{\hat{\rho}} & = \vec{d}_{eg} \left( a_{\ket{gg}\bra{eg}} + a_{\ket{ge}\bra{ee}} + a_{\ket{gg}\bra{ge}} + a_{\ket{eg}\bra{ee}} \right) \nonumber \\
	& \phantom{={}} +  \vec{d}_{ge}  \left( a_{\ket{eg}\bra{gg}} + a_{\ket{ee}\bra{ge}} + a_{\ket{ge}\bra{gg}} + a_{\ket{ee}\bra{eg}} \right)  \, .
\label{eq:MatrixElememtsDipoleMoment}
\end{align}
These elements thus indicate the strength of the classical coherence.

As a measure of the quantum coherence we will make use of the quantum discord and the concurrence. Let us first consider the quantum discord $D(B|A)$. It is defined as the difference of the quantum-mechanical mutual information $I(A\!:\!B)$ of two subsystems $A$ and $B$ and the classical correlations $J(B|A)$~\cite{Zurek(2000), Ollivier(2001), Modi(2012)}
\begin{equation}
D(B|A) = I(A\!:\!B)-J(B|A)\, ,
\label{eq:Discord}
\end{equation}
where the mutual information is given by
\begin{equation}
I(A\!:\!B)=S(A)+S(B)-S(AB) \, ,
\label{eq:MutualInformation}
\end{equation}
with $S(A)$ ($S(B)$) denoting the von Neumann entropy of system $A$ ($B$), and $S(AB)$ is the joint von Neumann entropy of the total system $AB$. 
The classical correlations in Eq.~(\ref{eq:Discord}) can be written as~\cite{Henderson(2001)} 
\begin{equation}
J(B|A)=\max_{\{\hat{E}_a\}} J(B|\{\hat{E}_a\})=\max_{\{\hat{E}_a\}}[S(B)-S(B|\{\hat{E}_a\})]\,,
\label{eq:ClassicalCorrelations}
\end{equation}
where
\begin{equation}
S(B|\{\hat{E}_a\}) = \sum_{a}p_a S(\hat{\rho}_{B|a})\,,
\label{eq:CondEntropy}
\end{equation}
with
\begin{equation}
\hat{\rho}_{B|a}=\frac{\text{Tr}_{A}[\hat{E}_a\hat{\rho}_{AB}]}{p_a}\,,
\label{eq:CondDensityMatrix}
\end{equation}
and $p_a = \text{Tr}[\hat{E}_a\hat{\rho}_{AB}]$ denoting the probability of the outcome $a$ of a measurement on subsystem $A$ described by the positive operator-valued measure $\hat{E}_a$. 
Note that for pure states $S(AB)=0$ and $J(B|A)=S(A)=S(B)$~\cite{Hall(2006)}, such that $D(B|A)=D(A|B)=S(A)=S(B)$.

In the case of pure states the quantum discord is directly related to the entanglement of the state. 
This is due to the fact that for pure states we find that $S(A)$ is equal to the entanglement of formation $E(\psi)$~\cite{Hill(1997),Wootters(1998),Modi(2012)}. The latter
can be written as~\cite{Wootters(1998)}
\begin{equation}
E(\psi)=h\left(\frac{1+\sqrt{1-c^2}}{2}\right)\,,
\end{equation}
where $c$ denotes the concurrence and $h$ the binary entropy function
\begin{equation}
h(x)=-x\log_2 x -(1-x)\log_2 (1-x)\,.
\end{equation}
Thus, for pure states, all quantum correlations measured by the quantum discord are due to $E(\psi)$. 

The concurrence is a positive and monotonous function, defined as \cite{Wootters(1998)}
\begin{align}
	c = \text{max}(0, \sqrt{\lambda_{1}}-\sqrt{\lambda_{2}}-\sqrt{\lambda_{3}}-\sqrt{\lambda_{4}}) \quad \in [0,1] \, ,
\label{eq:concurrence}
\end{align}
where $\lambda_{i}$ ($i=1,2,3,4$) are the eigenvalues of $\hat{\rho} \hat{\tilde{\rho}}$, in decreasing order, with $\hat{\tilde{\rho}}$ given by \cite{Wootters(1998)}
\begin{align}
	\hat{\tilde{\rho}} = \hat{\sigma}_{y} \otimes \hat{\sigma}_{y} \hat{\rho}^{*} \hat{\sigma}_{y} \otimes \hat{\sigma}_{y}\,.
	\label{eq:complexconjugate}
\end{align}
Here, the matrix $\hat{\rho}^{*}$ is the complex conjugate of $\hat{\rho}$ and $\hat{\sigma}_{y}$ denotes the Pauli spin matrix. As long as $c=0$, no entanglement is present. However, a nonvanishing concurrence, i.e., $c> 0$, indicates the presence of entanglement, where the strength of entanglement is given by the value of $c$, with a maximal entanglement obtained for $c=1$. 
For a maximally entangled pure state, we have ${D(B|A)=E(\psi)=c=1}$, whereas for a separable pure state we obtain $D(B|A)=E(\psi)=c=0$, i.e., all three quantities coincide.

\section{Symmetric Dicke state}
\label{sec:SymDickeState}

The first quantum state we want to investigate is the symmetric Dicke state of two two-level atoms with one excitation \cite{Dicke(1954)}
\begin{equation}
	\ket{\Psi} = \frac{1}{\sqrt{2}} \left( \ket{eg} + \ket{ge} \right)\,. 
\label{eq:DickeState}
\end{equation}
The corresponding density matrix can be written as
\begin{align}
	\hat{\rho}_{{}_\Psi}=\ket{\Psi}\bra{\Psi} =  \bordermatrix{
    ~ & gg & eg & ge & ee \cr
		gg & 0 & 0 & 0 & 0 \cr
    eg & 0 & \frac{1}{2} & \frac{1}{2} & 0 \cr
		ge & 0 & \frac{1}{2} & \frac{1}{2} & 0 \cr
		ee & 0 & 0 & 0 & 0  } \, .
		\label{eq:DickeMatrix}
\end{align}
As can be seen from Eq.~(\ref{eq:DickeMatrix}), the only non-vanishing entries of the density matrix are the four central elements. This immediately proves the non-separability of the state and, at the same time, demonstrates that the state $\ket{\Psi}$ gives rise to an interference pattern [cf. Eq.~(\ref{eq:Visibility})].

The intensity distribution produced by the state $\ket{\Psi}$ in the far field is well-known \cite{Wiegner(2011)PRA,Wiegner(2015)}. It calculates to [see Eqs.~(\ref{eq:G1Contributions})-(\ref{eq:InterferenceTerm})]
\begin{align}
	G^{(1)}_{\ket{\Psi}}(\delta) &= \bra{\Psi} \hat{E}^{(-)}(\delta) \hat{E}^{(+)}(\delta) \ket{\Psi} = 1 + \cos(\delta) \, ,
\label{eq:DickeIntensity}
\end{align} 
i.e., the visibility of the interference pattern is $100\%$. This is due to the fact that the system decays from a coherent superposition of two orthogonal states [cf. Eq.~(\ref{eq:DickeState})] into a single final state, i.e., $\ket{gg}$ \cite{Wiegner(2011)PRA}. The same result can also be obtained using the general expression of the visibility given in Eq.~(\ref{eq:Visibility}). Here again, with $a_{\ket{i}\bra{j}}=1/2$, $i,j\in\left\{eg,ge\right\}$, and $a_{\ket{ee}\bra{ee}}=0$, we obtain $\mathcal{V}_{\rho}=1$. Moreover, according to Eq.~(\ref{eq:DickeIntensity}), the maximum of the intensity pattern equals $2$, i.e., identical to the number of atoms $N$. Note that both, a maximal visibility of $100\%$ and the scaling of the central peak $\sim N$, are characteristic features of superradiance for systems containing a single excitation \cite{Wiegner(2011)PRA,Wiegner(2015)}.

Next, we prove that for the symmetric Dicke state $\ket{\Psi}$ the total dipole moment vanishes. To that aim, we calculate the expectation value of the total dipole moment [see Eq.~(\ref{eq:MatrixElememtsDipoleMoment})] yielding
\begin{equation}
 \bra{\Psi} \hat{\vec{D}} \ket{\Psi} = 0 \, .
\label{eq:Dickedipole}
\end{equation}
The result reaffirms the fact that there is no classical coherence between atoms prepared in the state $\ket{\Psi}$, a result which remains true for any kind of symmetric Dicke state, even for larger numbers of atoms or higher excitations \cite{Wiegner(2011)PRA}. The picture of classical antennas oscillating in phase leading to an enhanced emission of radiation, often employed in the context of Dicke superradiance, therefore cannot be used to explain the radiation emitted from such a system. 

Finally, we want to investigate whether the state $\ket{\Psi}$ possesses quantum coherence. To that aim, we calculate its concurrence. According to Eq.~(\ref{eq:concurrence}), we obtain 
\begin{align}
c_{\Psi}=1 \, .
\label{eq:Dickeconcurrence}
\end{align}
This shows that the state $\ket{\Psi}$ is maximally entangled, implying also maximal entanglement of formation as well as a maximal quantum discord (see Sec.~\ref{sec:SystemMeasures}). The outcomes of Eqs.~(\ref{eq:Dickedipole}) and (\ref{eq:Dickeconcurrence}) demonstrate that it is uniquely the quantum coherence in form of a maximal entanglement which is responsible for the superradiant interference pattern of Eq.~(\ref{eq:DickeIntensity}), not any kind of classical coherence in the form of a dipole moment. 

\section{Atomic product state of two atoms}
\label{sec:Dipoles}
Next we discuss the configuration of two identical synchronized two-level atoms, i.e., prepared in an identical coherent superposition of their ground and excited states $\ket{g}_{l}$ and $\ket{e}_{l}$, $l \in \left\{1, 2\right\}$, respectively. More specifically, we assume that the two atoms are prepared in $\ket{g}_{l}$ with probability $\alpha^{2} \in [0,1]$ and in $\ket{e}_{l}$ with probability $\beta^{2} \in [0,1]$, where $\alpha^{2} + \beta^{2} = 1$. The quantum state of the system reads \cite{Mandel(1983)}
\begin{align}
 \ket{\mathcal{P}} &= \left( \alpha \ket{g}_{1} + \beta \ket{e}_{1} \right) \otimes \left( \alpha \ket{g}_{2} + \beta \ket{e}_{2} \right) \nonumber \\
&  =  \alpha^{2} \ket{gg} + \alpha\beta \ket{eg} + \alpha\beta \ket{ge} + \beta^{2} \ket{ee} \, ,
	\label{eq:ProductState}
\end{align}
and the corresponding density matrix $\hat{\rho}_{{}_\mathcal{P}}$ takes the form
\begin{align}
	\hat{\rho}_{{}_\mathcal{P}}=\ket{\mathcal{P}}\bra{\mathcal{P}} =  \bordermatrix{
    ~ & gg & eg & ge & ee \cr
		gg & \alpha^{4} & \alpha^{3}\beta & \alpha^{3}\beta & \alpha^{2}\beta^{2} \cr
    eg & \alpha^{3}\beta & \alpha^{2}\beta^{2} & \alpha^{2}\beta^{2} & \alpha\beta^{3} \cr
		ge & \alpha^{3}\beta & \alpha^{2}\beta^{2} & \alpha^{2}\beta^{2} & \alpha\beta^{3} \cr
		ee & \alpha^{2}\beta^{2} & \alpha\beta^{3} & \alpha\beta^{3} & \beta^{4}  } \, .
	\label{eq:ProductMatrix}
\end{align}
From Eq.~(\ref{eq:ProductMatrix}), one can see that all entries of the matrix are non-vanishing in principle. We thus expect the system to display both, a modulated intensity distribution [cf. Eq.~(\ref{eq:Visibility})] as well as a classical dipole moment [cf. Eq.~(\ref{eq:MatrixElememtsDipoleMoment})]. 

The far field intensity distribution of two two-level atoms in a coherent superposition of ground and excited state has been investigated in much detail before (see, e.g., \cite{Mandel(1983)}). It calculates to 
\begin{align}
	G^{(1)}_{\ket{\mathcal{P}}}(\delta) &= \bra{\mathcal{P}} \hat{E}^{(-)}(\delta) \hat{E}^{(+)}(\delta) \ket{\mathcal{P}} = 2\beta^{2} \left[ 1 + \alpha^{2} \cos(\delta) \right] \, .
\label{eq:ProductIntensity}
\end{align}
For $\alpha=1$ and $\beta=0$, both atoms are prepared in the ground state, i.e., $\ket{\mathcal{P}} = \ket{gg}$,  in which case no photons are emitted from the atoms and the intensity vanishes identically, i.e., $G^{(1)}_{\ket{\mathcal{P}}}(\delta) = 0$. By contrast, for $\alpha=0$ and $\beta=1$, both atoms are prepared in the excited state, i.e., $\ket{\mathcal{P}} = \ket{ee}$, in which case the two independently decaying atoms end up in the two orthogonal (fully distinguishable) states $\ket{eg}$ and $\ket{ge}$. In this case, one obatins $G^{(1)}_{\ket{\mathcal{P}}}(\delta) = 2$, i.e., a constant, and again no modulation appears in the far field intensity distribution, i.e., $\mathcal{V}_{\rho}=0$. In both cases, the density matrix given in Eq.~(\ref{eq:ProductMatrix}) possesses only a single entry so that also the dipole moment vanishes.

\begin{figure}
	\centering
		\includegraphics[height=5.3cm]{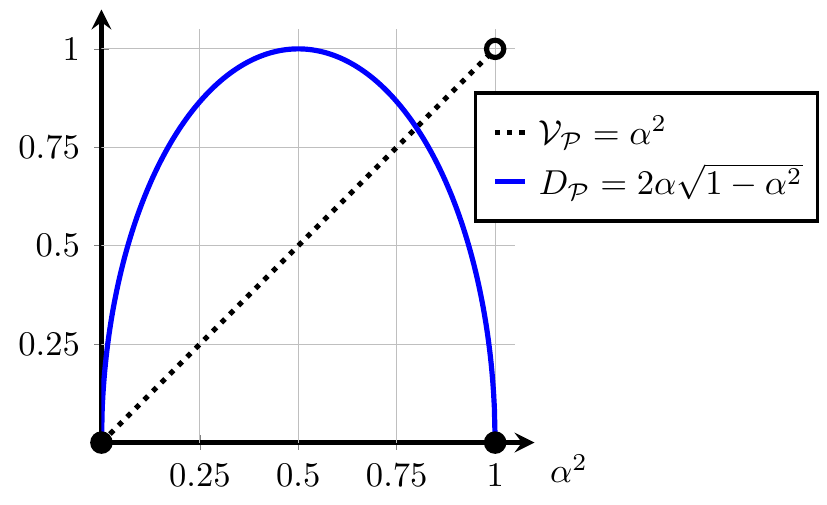}
	\caption{Visibility $\mathcal{V}_{\mathcal{P}}$ (dotted (black) line) [see Eq.~(\ref{eq:VisibilityP})] and dipole moment $D_{\mathcal{P}}$ (solid (blue) line) [see Eq.~(\ref{eq:DipoleMomentP})] for the state $\ket{\mathcal{P}}$ as a function of the population $\alpha^2$ of the ground states $\ket{g}_l$, $l\in \lbrace 1, 2\rbrace$, of the two atoms.
	\vspace{0.3cm}}
	\label{fig:Plot1}
\end{figure}

By contrast, for $0<\alpha < 1$ (and thus $0<\beta < 1$), $G^{(1)}_{\ket{\mathcal{P}}}(\delta)$ displays a modulation with a visibility [see Eq.~(\ref{eq:ProductIntensity})]
\begin{align}
	\mathcal{V}_{\mathcal{P}} = \alpha^{2} \, ,
\label{eq:VisibilityP}
\end{align}
whereas the contributing dipole moment reads [see Eq.~(\ref{eq:MatrixElememtsDipoleMoment})]
\begin{align}
	\bra{\mathcal{P}} \hat{\vec{D}} \ket{\mathcal{P}} & = 2\alpha \beta \left( \vec{d}_{ge} + \vec{d}_{eg} \right) \nonumber \\
	& = 2 \alpha \sqrt{1-\alpha^{2}} \left( \vec{d}_{ge} + \vec{d}_{eg} \right) \nonumber \\
	& \equiv D_{\mathcal{P}} \left( \vec{d}_{ge} + \vec{d}_{eg} \right) \, .
	\label{eq:DipoleMomentP}
\end{align}
In Fig.~\ref{fig:Plot1} the visibility $\mathcal{V}_{\mathcal{P}}$ and the dipole moment $D_{\mathcal{P}}$ are plotted with respect to $\alpha^{2}$. Obviously, $\mathcal{V}_{\mathcal{P}}$ scales linearly with the probability that the two atoms are prepared in the ground state $\alpha^{2}$ [cf. Eq.~(\ref{eq:ProductState})]. However, as can be seen from Eq.~(\ref{eq:DipoleMomentP}) and from Fig.~\ref{fig:Plot1}, the highest visibility (for $\alpha\rightarrow 1$) does not result from a strong dipole moment (which takes its maximal value for $\alpha^{2}=\beta^{2}=1/2$), but emerges from the fact that the $\ket{ee}\!\!\bra{ee}$-contribution (scaling with $\sim \beta^{4}$ and causing merely an additional background in the intensity distribution) vanishes faster than the interfering terms (scaling with $\sim \beta^{2}$) [see Eq.~(\ref{eq:ProductMatrix})]. Again, this is in contradiction to the classical expectation that a maximal dipole moment causes an interference pattern with maximal contrast \cite{Haroche(1982)}. Note that the effect of achieving a higher visibility for lower intensities is well-known and has been discussed, e.g., in Refs. \cite{Skornia(2001),Wolf(2016)}.

In order to prove that the intensity pattern produced by two synchronized atomic dipoles is of classical nature, independent of the visibility achieved, we calculate the concurrence $c$ of the density matrix $\hat{\rho}_{{}_\mathcal{P}}$ in Eq.~(\ref{eq:ProductMatrix}). From Eq.~(\ref{eq:concurrence}) we obtain
\begin{align}
c_{\mathcal{P}}= 0 \, ,
\end{align}
showing that there is no entanglement in the system and, hence, no presence of quantum coherence (see Sec.~\ref{sec:SystemMeasures}). Consequently, the interference pattern in the intensity distribution of two synchronously oscillating atoms results exclusively from classical coherence, i.e., from the dipole moments of the two atoms [cf. Eq.~(\ref{eq:DipoleMomentP})].

Note that for the investigated configuration, i.e., two atoms prepared in a coherent superposition of $\ket{g}_l$ and $\ket{e}_l$, $l\in \lbrace 1,2\rbrace$, the emitted intensity given in Eq.~(\ref{eq:ProductIntensity}) is of the same form as the intensity distribution of the maximally entangled symmetric Dicke state [cf. Eq.~(\ref{eq:DickeIntensity})], however with a reduced visibility ranging between $0 \leq \mathcal{V}_{\mathcal{P}} < 1$.

In the next section we investigate a third configuration, namely two two-level atoms being prepared in a Werner state. Here, we will show  that lower visibilities than $100\%$ can also be found from two two-level atoms in quantum states displaying only quantum coherence. In this way we demonstrate that the visibility of the interference patterns alone does not deliver any information about the type of coherence. 

\section{Werner state}
\label{sec:WernerState}

The statistical mixture of the maximally entangled Dicke state $\ket{\Psi}\bra{\Psi}$ [cf. Eq.~(\ref{eq:DickeState})] and the incoherent noise state $\hat{\mathbbm{1}}/4$ \cite{Yao(2015)} is the well-known Werner state \cite{Werner(1989)}
\begin{align}
	\hat{\rho}_{{}_\mathcal{W}} &= \frac{1-p}{4} \hat{\mathbbm{1}} + p \ket{\Psi}\bra{\Psi} \nonumber \\
	& = \bordermatrix{
    ~ & gg & eg & ge & ee \cr
		gg & \frac{1-p}{4} & 0 & 0 & 0 \cr
    eg & 0 & \frac{1+p}{4} & \frac{p}{2} & 0 \cr
		ge & 0 & \frac{p}{2} & \frac{1+p}{4} & 0 \cr
		ee & 0 & 0 & 0 & \frac{1-p}{4}  } ,
	\label{eq:StateWerner}
\end{align}
where $p \in [0,1]$ describes the relative weight between the two terms.
The spatial intensity distribution for the Werner state has been calculated before (see, e.g., \cite{Tang(2015)}) and can be shown to be
\begin{align}
	G^{(1)}_{\hat{\rho}_{{}_\mathcal{W}}}(\delta) = 1 + p \cos(\delta) \, ,
\label{eq:WernerIntensity}
\end{align}
with the visibility exclusively depending on the probability $p$ 
\begin{align}
\mathcal{V}_{\mathcal{W}} = p \, .
\label{eq:VisibilityW}
\end{align}
Note that for $p=0$ the interference pattern vanishes because the state is identical to the noise state, i.e., $\hat{\rho}_{{}_\mathcal{W}} = \hat{\mathbbm{1}}/4$. By contrast, for $p=1$ the interference becomes maximal since the state is identical to the symmetric Dicke state, i.e., $\hat{\rho}_{{}_\mathcal{W}} = \ket{\Psi}\bra{\Psi}$, displaying a visibility of $100\%$ [see Eq.~(\ref{eq:DickeIntensity})].  

From the entries of the density matrix of the Werner state [see Eq.~\eqref{eq:StateWerner}] one readily infers that the Werner state does not possess any dipole moment. Indeed, from Eqs.~(\ref{eq:MatrixElememtsDipoleMoment}) and (\ref{eq:StateWerner}), we obtain
\begin{equation}
\text{Tr}[\hat{\rho}_{{}_\mathcal{W}}\hat{\vec{D}}]=0 \, .
\end{equation}
This means that $\hat{\rho}_{{}_\mathcal{W}}$ does not contain any classical coherence.

\begin{figure}
	\centering
		\includegraphics[height=5.3cm]{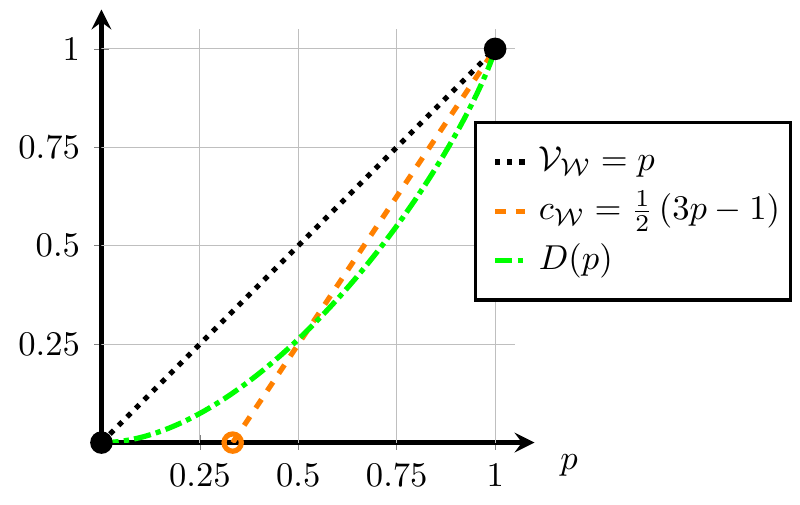}
	\caption{Visibility $\mathcal{V}_{\mathcal{W}}$ (dotted (black) line) [see Eq.~(\ref{eq:VisibilityW})], concurrence (dashed (orange) line) $c_{\mathcal{W}}$  [see Eq.~(\ref{eq:ConcurrenceW})], and quantum discord $D(p)$ (dotted-dashed (green) line) [see Eq.~(\ref{eq:discord})] as a function of the relative weight $p$ between the maximally entangled Dicke state $\ket{\Psi}\bra{\Psi}$ and the incoherent noise state $\hat{\mathbbm{1}}/4$ of the Werner state $\hat{\rho}_{\mathcal{W}}$.
	\vspace{0.3cm}}
	\label{fig:Plot2}
\end{figure}

By contrast, it is well-known that the Werner state possesses entanglement for $p> \frac{1}{3}$   
\cite{Tang(2015)}. Indeed, calculating the concurrence according to Eq.~(\ref{eq:VisibilityW}) one finds \cite{Tang(2015)}
\begin{align}
	c_{\mathcal{W}} =  \frac{1}{2} \left( 3p-1 \right) > 0  \quad \Rightarrow \quad p > \frac{1}{3} \, ,
\label{eq:ConcurrenceW}
\end{align}
whereas the quantum discord is given by~\cite{Tang(2015)}
\begin{align}
D(p)=&\;\frac{1-p}{4}\log_2(1-p)-\frac{1+p}{2}\log_2(1+p)\nonumber\\
&+\frac{1+3p}{4}\log_2(1+3p) \, ,
\label{eq:discord}
\end{align}
with $D(p) \geq 0$ independent of $p$.

In Fig.~\ref{fig:Plot2}, we display the visibility $\mathcal{V}_{\mathcal{W}}$, the concurrence $c_{\mathcal{W}}$, and the quantum discord $D(p)$ of the Werner state. As can be seen, the three quantities increase monotonously with $p$ up to their maximal value of 1~\cite{Tang(2015)}. This shows that for the Werner state stronger quantum correlations cause a more pronounced interference pattern, until, at ${p=1}$, the interference pattern displays maximal contrast and is governed by a maximal amount of quantum correlations, i.e., maximal entanglement (see Sec.~\ref{sec:SymDickeState}).

Fig.~\ref{fig:Plot2} proves that the Werner state can produce identical interference patterns as two radiating dipoles (see Sec.~\ref{sec:Dipoles}), however, stemming from a fundamentally different physical origin. While for the Werner state any interference pattern (for $p>0$) is solely due to quantum correlations and even entanglement (for ${p>1/3}$) and not caused by any dipole moment, i.e., classical coherence, the state of two synchronous dipoles [cf. Eqs.~(\ref{eq:ProductState}) and (\ref{eq:ProductMatrix})] produce an interference pattern solely due to their dipole moment, i.e., due to classical coherence, but never due to any kind of quantum correlations or entanglement. Furthermore, under no circumstances, two synchronous dipoles can reach a visibility of $100\%$. 

\section{Conclusion}
\label{sec:conclusion}

In conclusion, we demonstrated that the phenomenon of Dicke superradiance, exhibiting an interference pattern which displays enhanced and inhibited spontaneous emission, originates exclusively from quantum coherence. More specifically, we found that the symmetric Dicke state of two two-level atoms with a single excitation displays a visibility of the far field intensity distribution of $100\%$ but does not possess any dipole moment, the latter being a result which holds true for any symmetric Dicke state \cite{Wiegner(2011)PRA}. 
Consequently, Dicke superradiance cannot be viewed as a classical effect produced by synchronized atomic dipoles oscillating in phase. 
By contrast, investigating two separable synchronized atoms with identical dipole moments, we found that this configuration does not possess any quantum coherence but exclusively classical coherence. At the same time, the setup produces a similar interference pattern as two atoms in a single-excited Dicke state, however with a smaller visibility. Finally, we studied two atoms in a Werner state and showed that they generate an interference pattern of identical form as the two atomic dipoles. Since the Werner state does not possess any classical coherence but exhibits exclusively quantum coherence (for $p>0$) and even entanglement (for $p>1/3$) \cite{Tang(2015)} this proves that one can observe identical interference patterns, even with identical visibility, without knowing whether the interference stems from quantum coherence or from classical coherence. 

For determining the type of coherence and the origin of an interference pattern, we therefore have to resort to other measures than the visibility or the scaling of the intensity with the square of the number of atoms (i.e., $\sim N^2$), e.g., using the dipole moment as a measure for classical coherence, and the quantum discord or the concurrence as a measure for quantum coherence. We hope that the analysis presented in this paper helps discerning distinct phenomena which result from different types of coherence~\cite{Agarwal(1974)}. Insight into different types of coherence, potentially of different order, might be beneficial for both fundamental quantum optical investigations \cite{Ficek(2005),Zeilinger2012,Plenio2017,Plenio2021} as well as technological applications \cite{Classen2017,Richter2021}.

\section*{Acknowledgements}
D.B. gratefully acknowledges financial support by the Cusanuswerk, Bischöfliche Studienförderung. M.B. and J.v.Z. gratefully acknowledge funding and support by the International Max Planck Research School - Physics of Light. This work was funded by the Deutsche Forschungsgemeinschaft (DFG, German Research Foundation) -- Project-ID 429529648 -- TRR 306 QuCoLiMa ("Quantum Cooperativity of Light and Matter'').


\bibliography{Literature_coherence}

\end{document}